\documentclass[fleqn,10pt]{wlscirep}
\usepackage[utf8]{inputenc}
\usepackage[T1]{fontenc}
\usepackage{amsmath}
\usepackage{graphicx}
\usepackage{dcolumn}
\usepackage{bm}
\usepackage{epsfig}
\usepackage{float} 
\usepackage{color}
\usepackage{tikz}
\usepackage{subcaption}
\usepackage{comment}
\usepackage[normalem]{ulem}
\newcommand \beq{\begin{eqnarray}}
\newcommand \eeq{\end{eqnarray}}
\newcommand \bea{\begin{eqnarray}}
\newcommand \eea{\end{eqnarray}}
\newcommand \dvec{{\bf d}}
\newcommand \kvec{{\bf k}}

\newcommand\rvec{{\bf r}}

\newcommand\Rvec{{\bf R}}

\def\simge{\mathrel{%
       \rlap{\raise 0.511ex \hbox{$>$}}{\lower 0.511ex \hbox{$\sim$}}}}
\def\simle{\mathrel{
       \rlap{\raise 0.511ex \hbox{$<$}}{\lower 0.511ex \hbox{$\sim$}}}}

\def\beq {\begin{equation}}
\def\eeq {\end{equation}}
\def\w {\omega}

\def\bfk {\mathbf{k}}
\def\bfr {\mathbf{r}}

\newcommand{\bra}[1]{\langle #1|}
\newcommand{\ket}[1]{|#1\rangle}
\newcommand{\vo}{V$_2$O$_5$ }

\newcommand{\rv}{\mathbf{r}}
\newcommand{\kv}{\mathbf{k}}
\newcommand{\qv}{\mathbf{q}}

\newcommand{\rhot}{\tilde{\rho}}

\newcommand{\etal}{{\it et al.}}

\begin{document}

\title{
Delocalization of dark and bright excitons in flat-band materials and the optical properties of \vo}
\author[1,2,*]{Vitaly Gorelov}
\author[1,2]{Lucia Reining}
\author[3]{Martin Feneberg}
\author[3]{R\"{u}diger Goldhahn}
\author[4,5,6]{Andr\'e Schleife}
\author[7]{Walter R. L. Lambrecht}
\author[1,2,8]{Matteo Gatti} 
\affil[1]{LSI, CNRS, CEA/DRF/IRAMIS, \'Ecole Polytechnique, Institut Polytechnique de Paris, F-91120 Palaiseau, France}
\affil[2]{European Theoretical Spectroscopy Facility}
\affil[3]{Institute of Physics, Otto von Guericke University Magdeburg, Universit\"{a}tsplatz 2, 39106 Magdeburg, Germany}
\affil[4]{Department of Materials Science and Engineering, University of Illinois at Urbana-Champaign, Urbana, Illinois 61801, USA}
\affil[5]{Materials Research Laboratory, University of Illinois at Urbana-Champaign, Urbana, Illinois 61801, USA}
\affil[6]{National Center for Supercomputing Applications, University of Illinois at Urbana-Champaign, Urbana, Illinois 61801, USA}
\affil[7]{Department of Physics, Case Western Reserve University, Cleveland,
  OH-441-6-7079, USA}
\affil[8]{Synchrotron SOLEIL, L’Orme des Merisiers Saint-Aubin, BP 48 F-91192 Gif-sur-Yvette, France}
\affil[*]{corresponding author: vitaly.gorelov@polytechnique.edu}

\date{\today}

\begin{abstract}
The simplest picture of excitons in materials with atomic-like localization of electrons is that of Frenkel excitons, where electrons and holes stay close together, which is associated with a large binding energy. Here, using the example of the layered oxide \vo, we show how localized charge-transfer excitations combine to form excitons that also have a huge binding energy but, at the same time, a large electron-hole distance, and we explain this seemingly contradictory finding. The anisotropy of the exciton delocalization is determined by the local anisotropy of the structure, whereas the exciton extends orthogonally to the chains formed by the crystal structure. Moreover we show that the bright exciton goes together with a dark exciton of even larger binding energy and more pronounced anisotropy. 
These findings are obtained by combining first principles many-body perturbation theory calculations, ellipsometry experiments, and tight binding modelling, leading to very good agreement and a consistent picture. Our explanation is general and can be extended to other materials.   

\end{abstract}


\maketitle

\section{Introduction}

It is common belief that the exciton binding energy directly correlates with its spatial localisation  \cite{Kittel1986}.
The classical Wannier and Frenkel textbook models\cite{Knox1963,Bechstedt2014} describe, respectively, the limiting cases of 
weakly bound and delocalised excitons, typically found in ordinary semiconductors,
and tightly bound and localised excitons, observed in molecular or noble gas solids.
Excitons that stand out from this conventional expectation have always attracted considerable attention, as recently witnessed by  the surge of interest in two-dimensional (2D) materials \cite{Wang2018}. 
In fact, exciton binding energies in 2D materials are always large, irrespective of their spatial extent \cite{Cudazzo2016}, which can be rationalised as the consequence of reduced effective screening in low dimensions \cite{Cudazzo2011}.
Remarkably, there has   been a long standing controversy whether strongly bound charge transfer excitons, 
occurring for example in alkali halides, can be considered as Frenkel excitons \cite{Knox1963,Rohlfing2000,Abbamonte2008}: the origin of their spatial delocalisation has remained largely unexplained.
In low-dimensional transition-metal oxides the 
interplay between charge localisation, electron interactions, and reduced dimensionality gives rise 
to a wide variety of intriguing physical properties
while posing a great challenge for their theoretical interpretation \cite{Baeriswyl2004}. These complex materials therefore offer an ideal playground for unconventional exciton physics.

Vanadium pentoxide (V$_2$O$_5$) is a layered oxide that can be
 a prototype example of such a challenging material.
It is a ``ladder compound'' with a unique crystal structure: the layers consist of double zigzag oxygen and vanadium chains along the $y$ direction connected by a bridge oxygen forming  V-O-V rungs along the $x$ direction, and vanadyl oxygen atoms located below and above the vanadium atoms (see Fig.2(a)). The three-dimensional (3D) bulk
\vo is thus expected to exhibit also properties of 2D or even quasi-1D character\cite{Yosikawa97,Sucharitakul17}. \vo is supposed to be a simple  band insulator, with a relatively wide gap opening between filled O $2p$ bands and empty V $3d$ bands. The peculiar crystal structure is manifested in the weakly dispersing top-valence bands and, above the band gap, in two narrow bands separated from higher energy conduction bands\cite{Lambrecht1981,Eyert1998,Chakrabarti1999,Bhandari2015} (see Fig.4 and Supplementary Figure 1). These split-off bands are due to V $d_{xy}$ orbitals that have the smallest overlap with O $2p$ and hence the smallest bonding-antibonding splitting. As a result, they have a pronounced quasi-1D character, dispersing mainly along the chain  $y$ direction.
Unlike other vanadium oxides\cite{Imada1998},  pure bulk \vo remains non-magnetic and insulating at all temperatures.
Due to doping with alkali intercalation, or because of vanadyl oxygen vacancies, 
the split-off bands can be partially occupied, leading to  charge ordering and interesting 1D magnetic properties such as a possible 1D spin-Peierls transition at low temperature 
\cite{Isobe1996,Fujii1997,Smolinski1998,Mostovoy1999,Bhandari15b}.

The band structure appears hence to be well understood, but
an
unusually large discrepancy between the calculated  band gap and the experimental one has persisted for a long time.
The experimental band gap is widely
accepted to be $2.35$ eV, as determined from Tauc plots based on absorption spectra, with a slightly different onset depending on crystal direction\cite{Kenny1966}.
This value is larger, but only to a small extent, than the Kohn-Sham 1.7 eV band gap\cite{Eyert1998,Chakrabarti1999,Bhandari2015} calculated in the local 
density approximation (LDA) of  density functional theory (DFT). However, the state-of-the-art theoretical approach for band structures, namely, the GW approximation\cite{Hedin1965} to the self-energy of many-body perturbation theory, yielded a much larger gap\cite{Lany2013,vanSetten2017,Bhandari2015}, with the indirect minimum gap ranging from 3.5 to 4.0 eV depending on whether a perturbative G$_0$W$_0$ scheme \cite{Aryasetiawan1998} or quasiparticle (QP) self-consistent GW\cite{Schilfgaarde2006} (QSGW) was adopted (see Supplementary Table 1). 

Excitonic effects may explain the difference between the quasiparticle and optical gap, but have not been studied previously. Indeed, the Tauc plot analysis\cite{Kenny1966,Kang2013,Schneider2020} is not applicable when excitonic effects are present. Direct and inverse photoemission measurements\cite{Meyer2011} on \vo films, which  yield the quasiparticle gap and should be free from excitonic effects, have obtained the largest value for the gap, of 2.8 eV.
However, there is a large uncertainty on these data because of the limited resolution of these spectroscopies and the uncertainties concerning the position of the Fermi level. Moreover, it is still substantially smaller than the calculated quasiparticle gap.
The origin of this discrepancy has been a matter of debate for a long time. 
One hypothesis, in particular, was an inadequate description of screening in the GW calculations. First, since \vo is a strongly polar material, a source of additional screening could be the lattice polarization.
In Ref.~\citen{Bhandari2015}, on the basis of a model by Bechstedt {\it et al.}\cite{Bechstedt2005}, the self-energy was rescaled by a factor 0.38 improving the agreement of the onset energy with ellipsometry
data\cite{Parker1990}.  Later work showed that the effect was largely overestimated and that it should rather reduce the gap by an amount of the order of 0.2 eV, way too small to account for the gap overestimate 
\cite{BHANDARI2016,Radha2021}. While the formation of self-trapped electron polarons has been found to be important in V$_2$O$_5$
\cite{Scanlon2008,Watthaisong2019,Lappawat2021}, they are not expected to occur at the time scale of an optical absorption and can thus not explain the discrepancy. On the purely electronic side, the accuracy of QSGW calculations is limited by the fact that no electron-hole interaction effects are included in the calculation of $W$. Inclusion of this effect is estimated to reduce the self-energy by roughly 20\% \cite{Deguchi2016}, which is, however, much too little in order to restore agreement. Moreover, these modifications of the screening do not lead to significant improvement of the spectral shape.

On the other hand, there is also some discrepancy between different optical experiments. Reflectance and ellipsometry measurements\cite{Mokerov1976,Parker1990,Losurdo2000,Canillas2021,Kang2014} have obtained spectra for the  complex dielectric function $\epsilon(\w) = \epsilon_1(\w) + i \epsilon_2(\w)$ that agree on the main peak positions but differ in the onset energy of the first peak (see Supplementary Figure 2).
The discrepancies between different optical measurements may be due to the sample quality; indeed, 
it has been shown that the growth conditions and synthesis methods can significantly shift the absorption band edge\cite{Ostreng2012,Le2019,Wu2004,Scanlon2008,Kang2014}. 
Presence of strong excitonic effects hence remains the most promising hypothesis.  
Indeed, it is well known that exciton binding is stronger in lower dimensions. The topic of excitons in strongly anisotropic, in particular layered, materials meets increasing interest\cite{Xia2014,Fei2015,Huang2016,Li2017,Li2019,Das2019,Baldini2017}.  However, typical values for the exciton binding energy in such materials are rather of the order of 100 meV, whereas, here, a difference of more than 1 eV between experiment and theory has to be explained. On the other hand, 
optical properties of \vo have, to the best of our knowledge, always 
been calculated neglecting excitonic effects\cite{Atzkern2000,Bhandari2015,Szymanski2018},  
 and the material is too peculiar to rely on extrapolation from other compounds. In particular, the above materials mostly show  significant in-plane dispersion of the electronic bands, which is not true for V$_2$O$_5$.

At first sight, on the basis of the Wannier model, one would not expect strong exciton binding  in V$_2$O$_5$, with its anisotropic dielectric constant of the order of 5 \cite{Bhandari2014,Das2019,Clauws1976}. Indeed, for example several  transitions metal oxide perovskites have similar dielectric constants and an exciton binding energy that lies rather in the range of 100-200 meV\cite{Varrassi2021}.
However, the rich playground of possible excitations in materials that are more complex than common semiconductors or insulators has been explored only very partially. V$_2$O$_5$ may be representative for a class of materials, where the electronic charge has a  pronounced atomic-like character, a band ordering suggesting charge transfer excitations, and a non-trivial, anisotropic crystal structure. While significant many-body effects can be expected, it is impossible to predict their strength and nature without taking these various aspects into account.

In the present work we show that \vo is the prototype example for a material whose optical properties are dominated by very strong electron-hole binding, with excitons of noteworthy properties. We calculate optical spectra including excitonic effects from first principles by solving the Bethe-Salpeter equation (BSE)\cite{the-book}, and we validate the results with state-of-the-art ellipsometry experiments on high-quality single crystals. The very good agreement allows us to use the calculated data to highlight the intriguing nature of bright and dark excitons. 
In particular, we propose an exciton model that explains the counter-intuitive anisotropy and delocalisation of these charge transfer excitons, whose relevance goes well beyond the specific case of \vo. %

\section{Results}

\begin{figure*}[t]
\center
\begin{minipage}[b]{1.\columnwidth}
\includegraphics[width=\columnwidth]{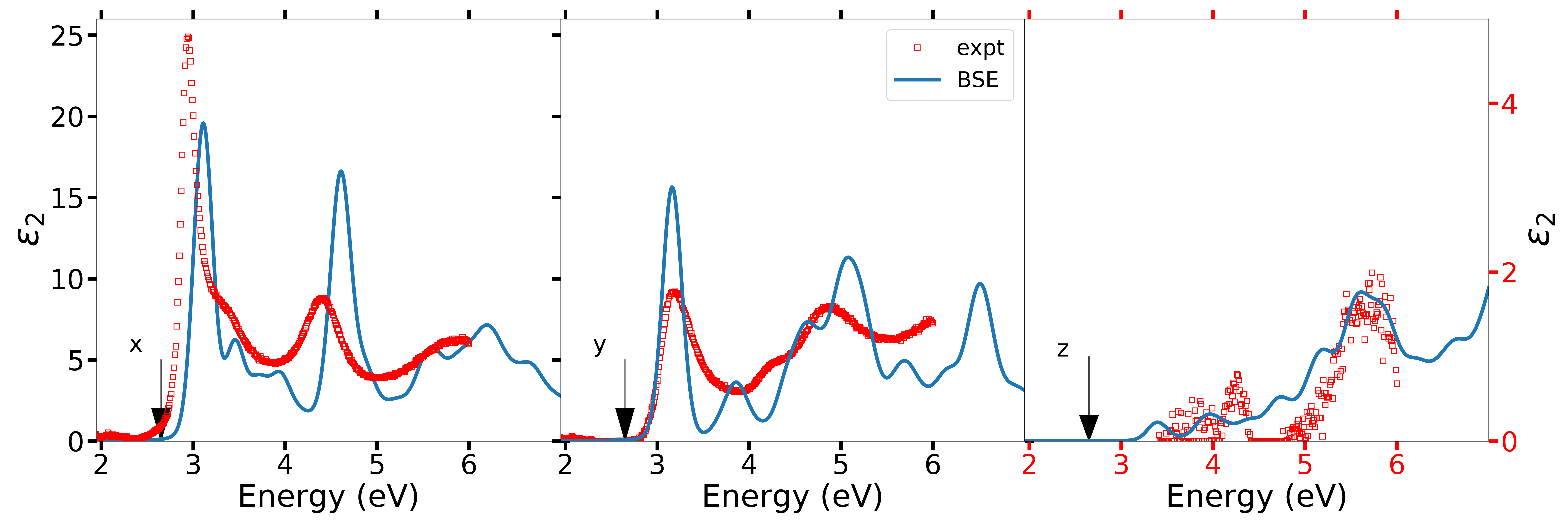}
\end{minipage}
\caption{\small{\textbf{Absorption spectra of \vo for $x$, $y$ and $z$ polarization}  (the \vo  layers are in the $xy$ plane). The experimental spectra obtained from ellipsometry  measurements  are compared with calculated BSE spectra in the three directions (note the change of intensity scale for the $z$ direction in the right panel). The vertical arrows mark the position of the dark excitons below the absorption onset.
\label{fig:BSE}}
}\end{figure*}

\subsection{Theoretical and experimental dielectric function}

We performed  state-of-the-art QSGW calculations to obtain an accurate QP band structure (see Fig.4 and Supplementary Figure 1).
Our results, with a direct gap at $\Gamma$ of 4.4 eV and an indirect minimum gap of 3.8 eV, agree to within 0.4 eV
with the previous QSGW  calculations\cite{Bhandari2015},  confirming the large discrepancy with respect to the
widely accepted experimental value for the gap near 2.35 eV \cite{Kenny1966} (see Supplementary Table 1). The subsequent solution of the electron-hole BSE
yields the absorption spectrum in Fig.1. For the in-plane polarizations, it shows a very pronounced peak at the onset, which is situated at about 3.1 eV, more than 1 eV below the direct QP gap. This shows that there is indeed a huge excitonic effect in this material. Our experimental results in Fig.1 confirm the \textit{ab initio} prediction: in the in-plane directions, they also show a very strong peak  at the onset around $\sim$ 3 eV, similar to, but stronger than in the experimental results of Mokerov \etal\cite{Mokerov1976} and Losurdo \etal\cite{Losurdo2000} (see Supplementary Figure 3). 
Most importantly, the broader peaks  found by Parker \etal \cite{Parker1990} below 3 eV are not confirmed by the present data. Our experiment and calculations also agree on the remaining features of the spectrum, in particular, 
the agreement is very good both in $x$ and $y$ direction. Along the chains, in the $y$ direction, the spectral shape is very well reproduced, and also smaller features such as the shoulder between 4 and 5 eV match. In $x$ direction, theory also correctly describes the 3-peak structure and the shoulder on the high energy side of the first peak. Absolute intensities are in qualitative agreement; note that on the experimental side, peak heights may show some variation according to the sample, and in particular to the surface roughness that enters the data analysis, and that broadening due to phonons is not included in the calculations. For the real part of the spectra see Supplementary Figure 4. The $z$ direction, i.e. out-of-plane polarization, is not directly accessible by our experiments due to the fact that the $z$ axis is aligned normal to the sample surface. Consequently, no high-quality experimental data is available for the out-of-plane polarization but the information that $\epsilon_2$ is of lower amplitude is reliable. On this basis, the qualitative agreement of theory an experiment for the $z$ direction is very satisfying.

The agreement between calculated and measured spectra resolves the long-standing apparent discrepancy between theory and experiment on the band gap of V$_2$O$_5$, indicating that the first peaks  at $\sim$ 3.1 eV in optical absorption are strongly bound excitons about 1.2 eV below the smallest (located at $Z$) direct QP gap of 4.3 eV.

\begin{figure*}[th]
\center
\begin{minipage}[b]{1.\columnwidth}
\includegraphics[width=\columnwidth]{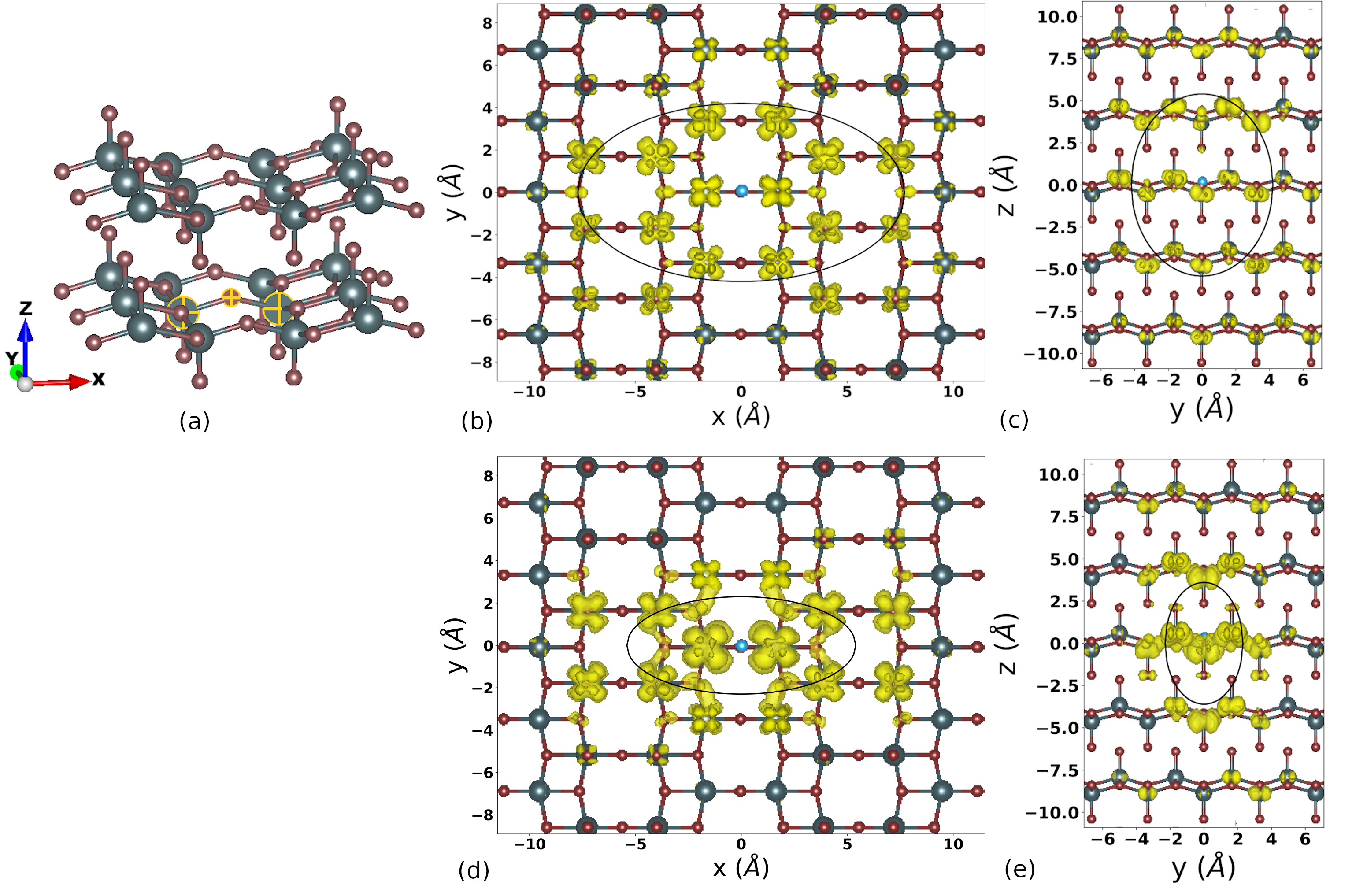}
\end{minipage}
\caption{\small{\textbf{Electron density distribution for dark and bright excitons.} (a) Crystal structure of \vo~. V atoms are grey, while O atoms are red. Selected atoms indicate the V-O-V rungs. (b)-(c) Electron density distribution $|\Psi_\lambda(\bfr_h^0,\bfr_e)|^2$ for the first exciton $\lambda$ that is bright in $x$ direction and (d)-(e) for the lowest-energy dark exciton. In both cases, the position of the hole $\bfr_h^0$ is fixed close to a bridge oxygen O$_b$ atom (light blue dot). Since it is displaced along a non-symmetric direction from the O$_b$ atom the resulting electronic distribution is also slightly asymmetric. (b) and (d) views in the plane of \vo layers (i.e., $xy$ planes) ; (c) and (e) views normal to \vo layers (i.e., $zy$ planes). Ellipses centered at the hole position represent the calculated exciton mean radius $\langle |r^{\alpha}_{\lambda}| \rangle$. The isosurface value for bright and dark excitons is chosen to be 1\% of the maximum electron density.
\label{fig:BSE_analys-1}}
}\end{figure*}

\subsection{Exciton extent and anisotropy}

We thus reveal a strong excitonic effect and strong anisotropy of the absorption spectrum in this material, which requires explanation. The
very good agreement between the calculations and experiment allows us to use the BSE results for the analysis.
The BSE can be formulated as a two-body Schr\"odinger equation in the basis of transitions between QSGW bands, with 
eigenvalues $E_{\lambda}$ and eigenvectors $\bar{A}^{vc{\bf k}}_{\lambda}$, where $v$ ($c$) stands for valence (conduction) bands and ${\bf k}$ is a wavevector in the first Brillouin zone (BZ) (see Methods). 
In the final spectrum, the coefficients $\bar{A}^{vc{\bf k}}_{\lambda}$ mix the QSGW transitions $vc{\bf k}$, and the contribution to the spectral intensity  for each excitonic transition $\lambda$ is $| \sum_{vc \kvec} \bar{A}^{vc \kvec}_{\lambda} \tilde{\rho}_{vc \kvec} |^2$, where $\tilde{\rho}_{vc \kvec} $ are the oscillator strengths of the QSGW transitions. This kind of calculation can therefore reveal both bright and dark excitons, the latter corresponding to $E_{\lambda}$ with vanishing $| \sum_{vc \kvec} \bar{A}^{vc \kvec}_{\lambda} \tilde{\rho}_{vc \kvec} |^2$. 
We find that indeed, the first transitions correspond to dark excitons, not visible in the spectrum in \textit{any} of the three polarization directions. They are found at an energy as low as $\sim$2.6 eV (see arrows in Fig.1), i.e., they have a binding energy as large as $E_b=1.6$ eV.

Inspection of the real-space  electron-hole correlation functions $|\Psi_\lambda(\bfr_h,\bfr_e)|^2$ (see Methods Eq. (\ref{eq:corfun})),
reveals further surprises. 
Fig.2(b)-(e) depict $|\Psi_\lambda|^2$ for the first dark exciton and the lowest energy exciton that is bright in the $x$ direction. 
With the hole fixed at position $\bfr_h^0$ next to a bridge O$_b$ atom (light blue dot), where the top-valence states contributing to these excitons are mostly located (see the analysis below with explanation of  Fig.2), the corresponding electron distribution $|\Psi_\lambda(\bfr_h^0,\bfr_e)|^2$ is represented as function of ${\bf r}_e$. For both excitons we find that  the electronic charge is entirely centered on V atoms, illustrating the O$_b$ $2p$ - V $d_{xy}$ charge transfer nature of the excitons, while the envelope functions globally have ellipsoidal shapes.
While the dark exciton (Fig.2(d),(e)) is more localised than the bright exciton (Fig.2(b),(c)), both excitons are delocalised across different \vo layers (see Figs.2(c) and 2(e)) and, to an even greater extent, within the \vo layers (see Figs. 2(b) and 2(d)), in stark contrast to what one would expect for such huge binding energies.
We have calculated the exciton mean radius over the supercell volume $\Omega$ for each direction $\alpha=\{x,y,z\}$ as  $\langle | r^{\alpha}_{\lambda}| \rangle = \frac1{\Omega} \int_{\Omega} d \rvec_e |r^{\alpha}_e-r^{0 \alpha}_h| |\Psi_{\lambda}(\mathbf{r}^0_h,\mathbf{r}_e)|^2$, finding an ellipsoid with axes (5.4, 2.3, 3.6)\AA{} and (7.7,  4.2, 5.4)\AA{}, which span more than a half unit cell and more than an entire unit cell, for the dark and bright exciton, respectively.
This result contradicts an extreme Frenkel picture of charge transfer excitons, where the exciton wavefunction would be perfectly localised within the V-O-V unit. A most counter-intuitive finding is the anisotropy, with especially the dark exciton extending in $x$ direction rather than along the chains, as the band structure would instead suggest. The reason will become clear in the following analysis, which highlights additional unusual facts.

\subsection{Analysis of the spectra}

\begin{figure*}[t]
\center
\begin{minipage}[b]{1.\columnwidth}
\includegraphics[width=\columnwidth]{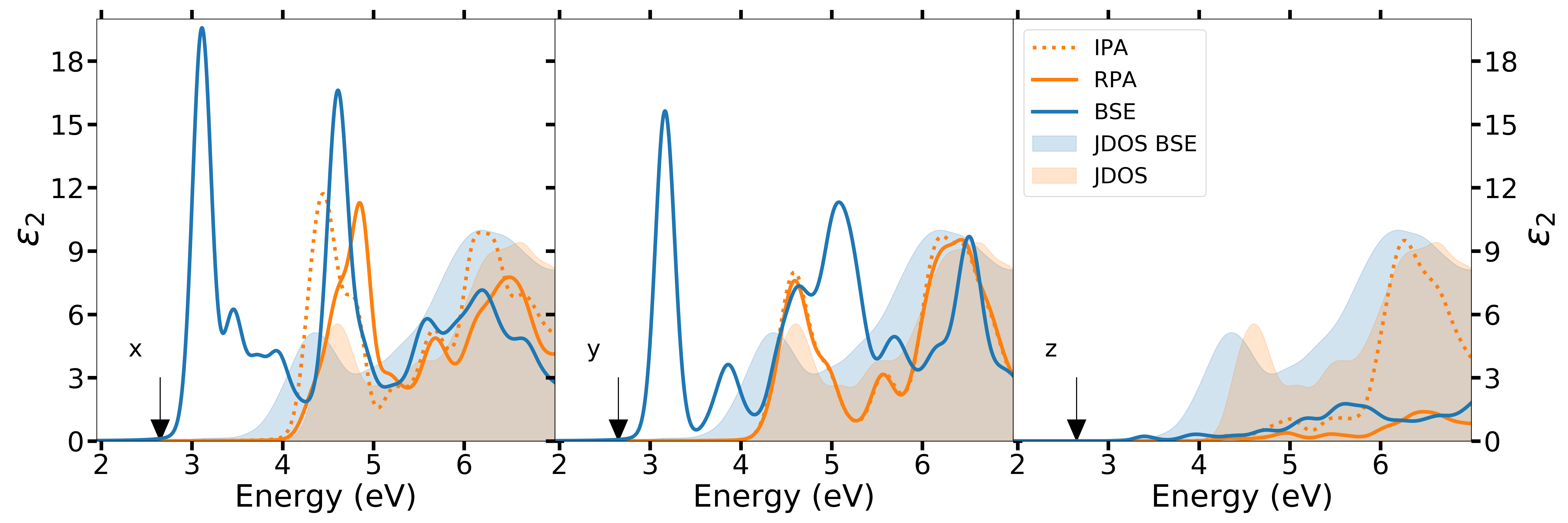}
\end{minipage}
\caption{\small{\textbf{Analysis of the absorption spectra of \vo for x, y, and z polarization.} The calculated JDOS and independent-particle (IPA) spectra  from the QSGW band structure are compared with the RPA spectra that include local field effects (LFE) and the BSE spectra that also include excitonic effects. The excitonic JDOS from the BSE, which is the density of excitonic states at zero momentum as a function of energy, is also shown for comparison. Discrete bound excitonic states inside the gap give rise to a JDOS with small intensity, not visible at the scale of the plot. The vertical arrows mark the position of the dark excitons below the absorption onset.
\label{fig:BSE_RPA}}
}\end{figure*}

The calculated spectra can be analysed as the combination of the joint density of states (JDOS), which reflects the density of electron-hole transition energies, and the dipole matrix elements weighting the transition intensities.
Let us start with the anisotropy of the spectra when comparing $z$ polarization to directions in the $xy$ plane. This anisotropy is extremely pronounced, although the bands have weak dispersion in all directions. 
Below 6 eV it is due to single-particle matrix elements, as one can see by comparing in Fig.3 the independent-particle spectrum to the JDOS stemming from the GW band structure.   
The matrix elements in the spectra enhance the first peak in the two in-plane directions $x$ and $y$ and strongly reduce it perpendicular to the planes, thus highlighting the layered character of the material. The spectrum is even more suppressed in $z$ direction, with a dramatic effect at higher energy, when, beyond the independent-particle approximation (IPA), crystal Local Field Effects (LFE) are taken into account in the Random Phase Approximation (RPA). 
LFE 
reflect the response on length scales of and below the crystal unit cell, and make the response function very sensitive to crystal anisotropies. The huge LFE perpendicular to the planes are reminiscent of the strong depolarization effects observed in finite systems, which indicates strong confinement in the planes \cite{Marinopoulos2002}.
Within the BSE, the effective unscreened electron-hole exchange interaction is responsible for LFE in the RPA, while the direct screened electron-hole attraction $-W$ describes excitonic effects beyond the RPA.
The LFE within the RPA can modify in principle the transition energies in the JDOS and the electron-hole wavefunctions in the matrix elements. 
Our RPA calculations have shown that, although transition energies are affected in principle, there is no visible change in the JDOS in practice. The suppression of the spectral weight is instead entirely due to changes in  the electron-hole wavefunctions, which changes the matrix elements.
Interestingly, we can now also clearly distinguish the behavior in $x$ and $y$ direction: in $y$ direction, along the chains, LFE are negligible, whereas perpendicular to the chains there is a sizable blueshift of oscillator strength and some suppression of intensity. This additional anisotropy, with electrons delocalized along the chains, gives the material a character between a 2D and a quasi 1D system. 

Changes of the JDOS between the IPA and the RPA 
are not analyzed usually.  Sometimes comparisons of JDOS are done between  IPA and full BSE calculations,  where it is commonly found that, besides the eventual appearance of bound states, modifications of the JDOS are small, even when spectra change dramatically\cite{Rohlfing2000,Yang2009}. This is true even in monolayers, where, besides the appearance of bound excitons, only minor (less than 0.1 eV) shifts of the JDOS are observed \cite{Ridolfi2018}.
This may be understood by considering 
the effective electron-hole interaction as a perturbation (see Methods).
In an extended system, first-order perturbation theory cannot change transition energies, which explains the stability of the JDOS, both in the RPA and full BSE calculations \cite{the-book}. Intensities, instead, may change already at first order, and the size of the changes is inversely proportional to energy differences between the unperturbed transitions. 
The spectral range shown in Fig.3 is dominated by transitions from the top-valence bands to the split-off conduction bands, which are dispersing very little.
Flat bands yield small energy differences between the transitions and can therefore explain huge changes of intensities, as observed here already in the RPA, even in first-order perturbation theory. 

 V$_2$O$_5$, however, behaves differently from other materials when the direct screened electron-hole interaction $-W$ is included in the BSE. Contrary to common findings, now also the JDOS undergoes significant changes all over the spectral range, as Fig.3 shows. In particular, there is 
an overall redshift of the order of half an eV. Note that the exciton binding energy is defined as $E_b=E_c-E_v -E_{\lambda}$, where $E_c-E_v$ is the energy of the lowest direct transition in QSGW, and $E_{\lambda}$ is the transition energy in the BSE. The substantial shift of the JDOS may require care for the definition of a binding energy extracted from experiment. This means that the excitonic effects are so strong that also a large spectral range above the continuum onset is dominated by significantly bound excitons. Instead, at higher energy, transitions that are more dispersing contribute to the spectra, resulting in a ~0.15eV red shift of the excitonic JDOS, which is in fact closer to other layered materials \cite{Ridolfi2018}.
 Finally, with respect to the BSE-JDOS, the BSE spectrum is governed by matrix element effects that are dramatic. They completely suppress the intensity of the two dark excitons at $\sim$2.6 eV, and create the huge oscillator strength of the bright excitons close to 3.1 eV. Such drastic matrix element effects are typically observed in low-dimensional systems \cite{Ridolfi2018}.

\begin{figure*}[th]
\center
\begin{minipage}[b]{1.\columnwidth}
\includegraphics[width=\columnwidth]{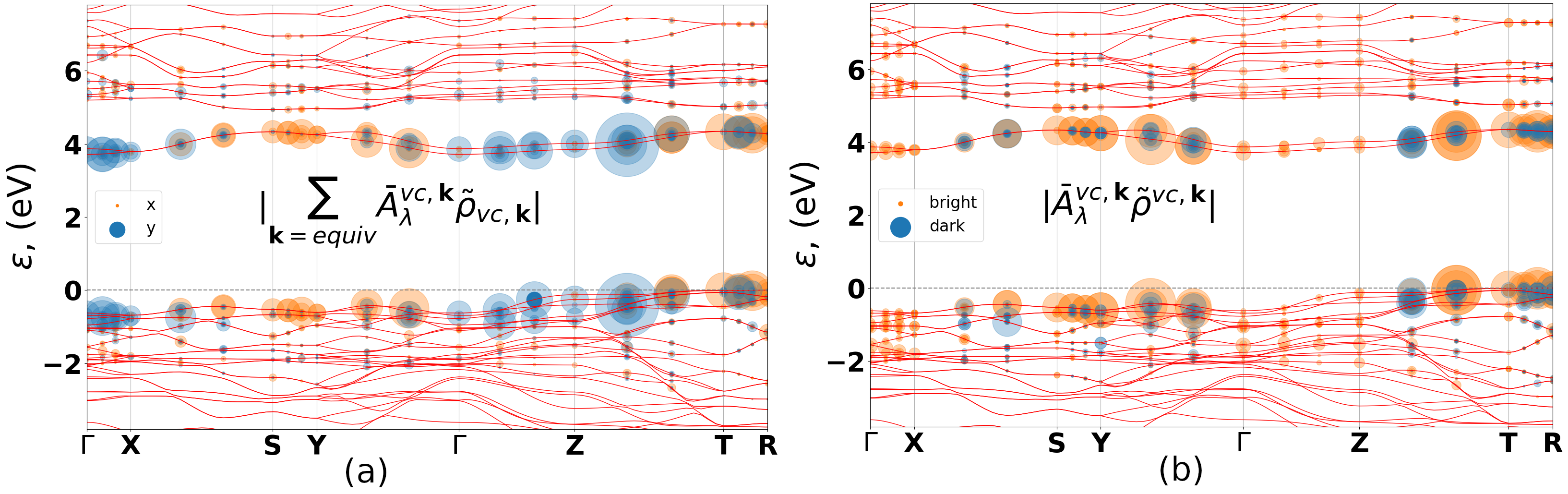}
\end{minipage}
\caption{\small{ \textbf{Contributions to the optical absorption intensity.} 
(a) For first bright excitons contributing to the main peaks in $x$ and $y$ directions (at 3.06 and 3.05 eV respectively). Size of the circles indicates $| \sum_{vc \kvec} \bar{A}^{vc \kvec}_{\lambda} \tilde{\rho}_{vc \kvec} |$ summed over equivalent $\kvec$ values. 
(b) For bright and dark exciton in $x$ direction. Size of the circles indicates $|\bar{A}^{vc \kvec}_{\lambda} \tilde{\rho}_{vc \kvec} |$  In both (a) and (b), transitions are vertical, so each transition links two circles of equal size, one in a valence and one in a conduction band, at the same ${\bf k}$ point. 
\label{fig:BSE_analys-2}}
}\end{figure*}

\subsection{Nature of the excitons}

To get a substantiated, though simple, explanation for our observations, the origin of the excitons must be analyzed. 
Fig.4(a) shows partial contributions to the intensity $|\sum_{{\bf k}_{equiv}}\bar{A}^{vc \kvec}_{\lambda} \tilde{\rho}_{vc \kvec} |$ projected onto the band structure, where the sum is performed  over ${\bf k}$ points that are equivalent by symmetry, for a bright exciton in $x$ direction and a bright exciton in $y$ direction at the absorption onset. 
The absolute value is indicated by the size of the circles. The circles always appear in pairs of equal size, one in a conduction and one in a valence band, at the same ${\bf k}$ point. 
The analysis shows that the character of excitons that are bright in $x$ and $y$ directions, respectively, is different.  The excitons at the absorption onset that are bright in $x$ direction are dark along $y$, and vice versa. The darkness of the excitons in certain directions is   
due to the negligibly small  oscillator strengths of the single particle transitions that are mixed to yield the excitonic transitions.  
In $x$ direction, all peaks are coming  predominantly from transitions around the S-Y-$\Gamma$ and Z-T regions in the Brillouin zone. In the Z-T region the transitions are from the bridge O$_b$ $p_z$ to the split-off V $d_{xy}$ states \cite{Bhandari2015}. Excitons that are bright in $y$ directions are made mainly of  transitions around $\Gamma$-X and $\Gamma$-Z-T between the top valence and the split-off conduction bands. 
The higher energy excitons, e.g. in $x$ direction at 4.55 eV, instead also mix transitions from lower lying valence and higher conduction bands (see Supplementary Figure 5).

The importance of the split-off conduction  bands for the spectrum in the low energy range explains many of the observations, including the fact that the calculations are quite delicate. In particular it is interesting to note that approximating the QSGW corrections to the LDA band structure by a rigid shift, as it is often done with the ``scissor operator approximation'',  would not be adequate here. Using a scissor, the highest valence and lowest conduction bands are not matching the QSGW band structure sufficiently well, leading to a splitting of the first two excitonic peaks into three peaks of equal intensity. This is shown in the Supplementary Figure 6. Indeed, although the discrepancies between QSGW and the scissor appear to be small in the band structure, the fact that excitonic effects in this material are so much due to mixing of transitions close in energy makes the result exceptionally sensitive to these changes.

The analysis allows us to set up a minimal tight-binding (TB) model (for details, see Supplementary Note 7) that should describe the exciton that is bright in $\hat{\bf x}$ direction: with the hole localized mainly on the bridge oxygen $p_z$, and the main contribution to the electron stemming from neighbouring V $d_{xy}$, neglecting the slight buckling of V-O-V rungs in $\hat{\bf z}$ direction, the TB wavefunctions of the corresponding valence and conduction band read
\begin{eqnarray}
&&\varphi_{v\bfk}(\bfr)=\frac{1}{\sqrt{N}}\sum_{\bf R} e^{i\bfk{\bf R}}\Phi_v(\bfr-{\bf R})\nonumber\\
&&\varphi_{c\bfk}(\bfr)=\frac{1}{\sqrt{N}}\sum_{\bf R} \Big (e^{i\bfk({\bf R}-{\bf d})}\Phi_c(\rvec-{\bf R}+{\bf d}) + e^{i\bfk({\bf R}+{\bf d})}\Phi_c(\bfr-{\bf R}+\sigma_x{\bf d})\Big )\,,
\end{eqnarray} 
where $N$ is the number of the unit cells that are identified by lattice vectors $\Rvec$, $\Phi$ are atomic orbitals, and ${\bf d}$ is the distance between the O$_b$, located at the origin, and the next neighbour V atoms that are related by a mirror plane $\sigma_x$ perpendicular to $\hat {\bf x}$.
The odd character of the matrix elements of $\nabla$, $\tilde{\rho}_{vc k_x}=-\tilde{\rho}_{vc k_x}$, becomes immediately clear. Note that the symmetry $\kvec \to -\kvec$ is restored if the second V-O-V rung in the unit cell is taken into account. The matrix elements of the bare exchange and direct screened electron-hole interactions are evaluated as prescribed by  Eqs. \eqref{eq:mv} and \eqref{eq:mW} in Methods. Taking into account a single pair of bands and neglecting the overlap between orbitals on different sites, this yields $v_{\bfk\bfk'}\approx \frac{2v}N \cos [(k_x-k_x')d_x] e^{i(k_z-k_z') d_z}$ with $v \equiv 2\sum_{\Rvec}\int d\rvec d\rvec'\,\Phi_v(\rvec)\Phi^*_v(\rvec') \bar{v}_c(\rvec-\rvec'+\Rvec)\Phi_c(\rvec'+\dvec)\Phi^*_c(\rvec+\dvec)$ for the electron-hole exchange, and, supposing the screened Coulomb interaction $W(\bfr,\bfr')$ to be short ranged and only dependent on the distance $|\bfr-\bfr'|$, one obtains for the direct electron-hole interaction  $-W_{\bfk\bfk'}\approx -\frac{w}{N}\, {\rm cos} ([k_x-k_x']d_x)e^{i(k_z-k_z') d_z}$ with $w \equiv 2\int d\rvec d\rvec'\,|\Phi_c(\rvec+\dvec)|^2 W(\rvec-\rvec')|\Phi_v(\rvec')|^2$ determined by the next-neighbour screened interaction. The resulting BSE can be solved analytically. Besides the continuum solutions, there is a solution $A^{\bfk}\propto {\rm cos} (k_xd_x) e^{ik_z d_z}$  that is even in $k_x$, with a binding energy of the order of $w-2v$, and an odd solution $A^{\bfk}\propto {\rm sin} (k_xd)e^{ik_z d_z}$ with reduced binding energy. Since the matrix elements are odd, the solution with smaller binding energy must be the bright exciton. The very strongly bound dark exciton, instead, is predicted to stem from the same single-particle transitions but with even $A^{\bfk}$. This is confirmed by the analysis of the \textit{ab initio} results in Fig.4(b): 
as for the bright excitons, the single-particle transitions $\tilde{\rho}_{vc \kvec}$ that dominate the dark exciton are \textit{not} dipole forbidden, so each contribution to the oscillator strength separately is not zero, but of the same order as for the bright excitons, in contrast to other materials with interesting strongly bound dark excitons, such as, e.g., hBN, CrI$_3$, and transition metal dichalcogenides, where dark excitons are made of dipole forbidden matrix elements \cite{Wirtz2008,Wu2019,Zhang2015,Malic2018}. 
As we have verified in the \textit{ab initio} calculations, in \vo the oscillator strengths are odd functions of $\kvec$, while the coefficients $\bar{A}^{vc{\bf k}}_{\lambda}$ are even functions in $\bfk$ . Therefore, when the contributions from 
$\kvec$ and $-\kvec$ points are summed together, they cancel each other and make the exciton dark, as predicted by the model.
Such a robust dark exciton can play a key role for many disexcitation processes\cite{Feierabend2017,Zhang2017,Kusaba2021,Park2017}, such as phonon-assisted luminescence and light emission. 

The model further allows us to understand the electron-hole correlation, using the
 $A^\bfk$ and tight-binding wavefunctions to yield the squared electron-hole wavefunction 
\begin{eqnarray} 
\label{eq:extensions}
    |\Psi(0,\bfr_e)|^2 &=& |\sum_\bfk A^\bfk\varphi^*_{v\bfk}(0)\varphi_{c\bfk}(\bfr_e)|^2 \nonumber\\
    &\propto&  \Big (|\Phi_c(\bfr_e+{\bf d})|^2 + |\Phi_c(\bfr_e+\sigma_x {\bf d})|^2\Big )\Big (1\pm\frac{\sin(2\pi d_x/a)}{4\pi d_x/a}\Big )^2 \\ \nonumber
    &&+
    \delta_{R_y,0}\delta_{R_z,0}\frac{ \sin^2(2\pi d_x/a)}{4\pi^2/a^2} \sum_{R_x\neq 0}\Big (\frac{ |\Phi_c(\bfr_e-\hat{\bf x}R_x+\dvec)|^2}{(R_x-2d_x)^2}  +\frac{|\Phi_c(\bfr_e-\hat {\bf x}R_x+\sigma_x \dvec)|^2}{(R_x+2d_x)^2}  \Big ) \Bigg],
\end{eqnarray}
where $a$ is the lattice parameter in $\hat{\bf x}$ direction, the upper (lower) sign in $\pm$ refers to the dark (bright) exciton, respectively, and the hole  has been placed on the bridge oxygen at $\bfr_h=0$.
The model predicts that the excitons are
 localized in $\hat{\bf y}$ and $\hat{\bf z}$ directions, but decay only quadratically with the distance in $\hat{\bf x}$ direction due to a contribution that depends on the O-V distance $d_x$. For $d_x\to 0$ this longer-range contribution would approach zero, and the picture would reduce to the Frenkel exciton. For $d_x\neq 0$, instead, the first term in \eqref{eq:extensions} yields the delocalization over the charge transfer unit that one would expect from the Frenkel picture, whereas the second term switches on the much larger $1/R_x^2$ extension. Indeed, although the model takes only the  next neighbour transition into account, it correctly predicts that the exciton is significantly more extended than the next neighbour distance. In other words, the delocalization in $\hat{\bf x}$ direction is due to the charge transfer nature of the main underlying single-particle transition, and not to the long range of the Coulomb interaction. Interestingly, it is also independent of the interaction strength , whereas the binding energy depends linearly on the interaction strength. This limited role of the interaction strength is a general characteristic of excitons stemming from isolated groups of weakly dispersing transitions (see Supplementary Note 8 for a proof), independently of the present tight binding model. The tight binding analysis, instead, shows that the local environment of the hole, with the V-O-V bridge, determines the shape of the exciton, rather than the atomic chains, which explains the counter-intuitive anisotropy of the wavefunction.  
 Of course, the model is simple, but it is validated by predicting further interesting details that are indeed observed in the \textit{ab initio} results, in particular, the weaker intensity of the density on the next-neighbour V atom in the bright, as compared to the dark exciton, or the ratio between the dark and bright electron-hole binding energy (see Supplementary Note 7).
 Because of this difference in density, looking at Figs.2(b) and 2(d) one would expect higher intensity for the lowest bound state, which instead, as predicted by the model and found in the  \textit{ab initio} calculations, is dark exclusively because of the symmetry of the mixing coefficients.

\section{Discussion}
In conclusion, we have shown that excitons in flat band materials may have intriguing properties which contradict textbook expectations. Flat bands correspond to localized electronic states, and as one would expect, the exciton binding energy is in general very large\cite{Radha2021}.  However, when the single-particle excitations that are mixed to form an exciton are charge transfer excitations with a mirror symmetry, the exciton wavefunction delocalizes. This delocalization is neither isotropic nor does it follow the overall anisotropy of the crystal structure, but it shows an anisotropy governed by the local motif of the charge transfer unit. Our work predicts in particular that materials with such a local geometry, where excitations take place from a central atom to two equivalent, symmetrically placed, neighbouring atoms, will show a bound exciton that may be more or less bright, according to the character and the overlap of the involved orbitals. On top of this, one will also find an even more strongly bound exciton that will always be dark because of destructive interference. While it may not be detected in optical measurements, its presence can be inferred from that of a strongly bound bright exciton and the appropriate local geometry. With the hole on the central atom, the electron distribution of both the dark and bright excitons has a large extension perpendicular to the mirror plane. 

A very good material to illustrate these effects is V$_2$O$_5$. As we have shown using first principles calculations and ellipsometry measurements, its optical properties are dominated by huge anisotropy and excitonic effects. 
Strongly bound dark and bright excitons are found in the gap, consistently with our  prediction, and even the excitation energies in the continuum are shifted significantly, contrary to usual observations. 
In the lowest dark and bright excitons, the electron density for a hole on a bridge oxygen  extends perpendicular to the atomic chains along which electrons disperse. Our explanation of the absorption spectra, based on a consistent picture emerging from state-of-the-art \textit{ab initio} calculations within many-body perturbation theory, a simple tight-binding model, and our  ellipsometry measurements, solves the long-standing puzzle of the optical properties of V$_2$O$_5$, and may serve as a guideline for the understanding of electronic excitations in other charge transfer insulators.
Moreover, our work illustrates that one cannot simply extrapolate knowledge gained from textbook semiconductors and insulators, such as the link between the dielectric constant, the electron-hole distance and the binding energy, to the advanced materials of current interest. This gives new freedom to engineering properties of materials; for example, we have shown that for materials where excitons are dominated by isolated groups of weakly dispersing excitations, the screening of the electron-hole interaction has a direct impact on the exciton binding energy but not on the electron-hole wavefunction. Therefore, by modifying the screening\cite{RiisJensen2020b} one can tune the absorption onset while keeping intact the spectral shape with its very strong intensity at the onset. Altogether, the present work may give guidelines to identify or combine materials where excitons exhibit properties tailored for specific needs.

\section{Methods}

\subsection{Optical properties from the Bethe-Salpeter equation}

We solved the electron-hole Bethe-Salpeter equation (BSE) 
using the QSGW QP eigenvalues and wavefunctions, and the screened direct electron-hole interaction given by the statically screened $W$ evaluated in the Random Phase Approximation (RPA) using QSGW ingredients.
The BSE is an in principle exact equation for the two-particle correlation function\cite{the-book}.
In the GW approximation\cite{Hedin1965}, the BSE with a statically screened Coulomb interaction $W$  can be cast as
an effective two-particle Schr\"odinger equation for the two-particle correlation function  of the electron-hole (e-h) pair\cite{Hanke1979,Albrecht1998,Rohlfing2000,Onida2002}:  $H_{\rm exc} \Psi_\lambda(\bfr_h,\bfr_e) = E_\lambda \Psi_\lambda(\bfr_h,\bfr_e)$.
This equation is usually expressed in a basis of pairs of orbitals. In gapped systems at zero temperature, only pairs of an occupied  $v{\bf k}$ and an unoccupied  $c{\bf k}$ orbitals contribute to an absorption spectrum, so the pair corresponds to a direct transition 
$|v c {\bf k}\rangle$.  
In this basis the resonant part of the Hamiltonian reads
$ \bra{vc{\bf k}} H_{\rm exc} \ket{v'c'{\bf k}'}=E_{vc{\bf k}} + \bra{vc{\bf k}} \bar{v}_c-W \ket{v'c'{\bf k}'}$.
Here the energy $E_{vc{\bf k}}$ is the difference between an unoccupied and an occupied quasiparticle (QP) state, calculated in the QSGW. The effective electron-hole exchange interaction $\bar v_c$ is given by the microscopic components of the bare Coulomb interaction and is responsible for LFE.  
The screened Coulomb interaction $-W$ represents the direct electron-hole interaction and accounts for excitonic effects.
If it is neglected, the BSE yields the RPA. The matrix elements in terms of  QP orbitals $\varphi$ read

\begin{equation}
\bra{t} \bar{v}_c \ket{t'}\equiv  2   \int d{\bf r}d{\bf r}'\,\varphi^*_{c{\bf k}}({\bf r})\varphi_{v{\bf k}}({\bf r})\bar{v}_c({\bf r},{\bf r}')\varphi_{c'{\bf k}'}({\bf r}')\varphi_{v'{\bf k}'}^*({\bf r}')
\label{eq:mv}
\end{equation}
and
\begin{equation}
\bra{t} W \ket{t'}\equiv \int d{\bf r}d{\bf r}'\, \varphi_{c{\bf k}}^*({\bf r})\varphi_{c'{\bf k}'}({\bf r}) W({\bf r},{\bf r}')\varphi_{v{\bf k}}({\bf r}')\varphi^*_{v'{\bf k}'}({\bf r}')  \,,
\label{eq:mW}
\end{equation}
where $t\equiv v,c,{\bf k}$. The two-particle correlation function is
 \beq
 \label{eq:corfun}
 \Psi_\lambda(\bfr_h,\bfr_e) = \sum_{vc{\bf k}} \bar A_\lambda^{vc{\bf k}} \varphi^*_{v{\bf k}}({\bf r}_h)\varphi_{c{\bf k}}({\bf r}_e).
 \eeq

In the Tamm-Dancoff approximation, 
which neglects the coupling between resonant and antiresonant transitions and is usually a good approximation for absorption spectra of solids (we have verified that this is the case also for \vo),
the macroscopic dielectric function is obtained from the BSE as: 
\beq
\epsilon_M(\w) =  1 - \lim_{\qv\to 0}\frac{8\pi}{N_k\Omega_0 q^2} \sum_\lambda \frac{\left|\sum_{t} \bar A_\lambda^{t} \tilde{\rho}_{t}(\qv) \right|^2 }{\w- E_\lambda + i\eta},
\label{spectrumBSE2}
\eeq
with $\Omega_0$ the unit cell volume and $N_{\bf k}$  the number of ${\bf k}$ points, and where the 
oscillator strengths are
$\rhot_{t}(\qv)= \int \varphi^*_{v\kv-\qv}(\rv) e^{-i\qv\cdot\rv}\varphi_{c\kv}(\rv) d\rv$.
Each exciton $\lambda$ contributes with strength $\left|\sum_{t} \bar A_\lambda^{t} \tilde{\rho}_{t}\right|^2$ to the absorption spectrum $\epsilon_2(\w) \equiv \textrm{Im} \epsilon_M(\w)$. If it is negligibly small, the exciton is said to be dark. Setting all  exciton strengths to 1 gives the excitonic JDOS: $\frac{2}{N_k\Omega_0 \w^2} \sum_\lambda \delta(\w- E_\lambda)$. The RPA or the independent-particle JDOS are obtained analogously by neglecting either only $-W$ or also $\bar v_c$.

\textbf{Analysis in terms of perturbation theory:}
Perturbation theory may help to understand the impact of flat bands on excitonic effects.
To zero order, a transition $\lambda$ corresponds to a band-to band transition $v_{\lambda}{\bf k}_{\lambda}\to c_{\lambda}{\bf k}_{\lambda}$, so $E^{(0)}_{\lambda}= E_{c_{\lambda}{\bf k}_{\lambda}}-E_{v_{\lambda}{\bf k}_{\lambda}}$ and $\bar A_{\lambda}^{vc{\bf k},(0)}=\delta_{vv_{\lambda}}\delta_{cc_{\lambda}}\delta_{{\bf k}{\bf k}_{\lambda}}$. First-order changes of the transition energies are given by the diagonal elements  $\langle t_{\lambda}|\bar v_c|t_{\lambda}\rangle$ and $\langle t_{\lambda}| W|t_{\lambda}\rangle$, which are vanishing in an infinite system because of the normalization of the wavefunctions over the crystal volume \cite{the-book}, as one can also see directly in the tight binding matrix elements in the Supplementary Note 7.

To first order, the corrections to intensities  $\sum_t \bar A_{\lambda}^{t}\tilde \rho_t$ 
are
\begin{equation}
   \sum_t \bar A_{\lambda}^{t,(1)}\tilde \rho_t
   = 
   \sum_{t\neq t_{\lambda}}
   \frac{\langle t|\bar{v}_c|t_{\lambda}\rangle-\langle t|W|t_{\lambda}\rangle}{E^{(0)}_{t_{\lambda}}-E^{(0)}_t} \tilde \rho_t\,.
\end{equation}

This expression shows that the sum over $t$ makes the result converge to a non-vanishing correction even for an infinite system with dense Brillouin  zone sampling. It also shows that changes in intensities are large when band-to-band transitions lie close in energy, because the differences appear in the denominator. Finally, for flat bands, where the differences can be very small even at distant ${\bf k}$-points, small changes in band energies can significantly alter the results.

\subsection{Computational details}

We adopted the $Pmmn$ orthorhombic crystal structure of \vo with 14 atoms per unit cell and the experimental lattice parameters\cite{Enjalbert1986} $a=11.512$ \AA{} , $b=3.564$ \AA{}  and $c=4.368$ \AA{}.  
We used norm-conserving Troullier-Martins\cite{Troullier1991}  pseudopotentials, including  $3s$ and $3p$ semicore states for vanadium (total of 112 electrons), which have been already validated in previous studies on vanadates\cite{Papalazarou2009,Gatti2015}. 
The LDA ground-state calculation converged with an energy cutoff of 100 Hartree and $4\times4\times4$ $\bfk$-point grid.
In the GW calculations, done using the Godby-Needs plasmon pole model\cite{Godby1989} ($\omega_p$ = 26 eV) and validated with the contour deformation integration technique\cite{Lebegue2003}, the dielectric function was computed using $6\times6\times6$ $k$-point grid and 350 bands and had a size of 4.9 Hartree, while the self-energy required 700 bands and 52 Hartree cutoff energy (see Supplementary Figure 2). Within the QSGW scheme\cite{Schilfgaarde2006} all O $2p$ bands (i.e., 30 topmost occupied bands)
and 22 empty bands were calculated self-consistently.
The BSE Hamiltonian was built using QSGW QP energies, wavefunctions and statically screened $W$
 with a $6\times6\times6$ $k$-point grid, and 15 valence and 16 conduction bands (see Supplementary Note 6 for more details). A 0.1eV Gaussian broadening was applied to the resulting spectra.
 LDA and GW calculations were carried out with ABINIT \cite{Gonze2005}, while the EXC code\cite{EXCcode} was used for BSE calculations.

\subsection{Experimental details}

Two V$_2$O$_5$ samples grown by floating zone melting\cite{Jachmann2005} with the surface normal parallel to $z$ ($\parallel c$) were investigated by spectroscopic ellipsometry in two different geometries, once with the $x$ direction ($\parallel a$) in the plane of incidence and once perpendicular to the plane of incidence. Both samples were measured under several angles of incidence ($\Phi = 50^\circ$, $60^\circ$, and $70^\circ$). The ellipsometric parameters $\Psi$ and $\Delta$ were determined for photon energies between 0.5 and 6 eV using generalized ellipsometry.  Spectral resolution was set to 1.3nm corresponding to about 10meV at 3eV. More details about the experimental procedure can be found e.g. in Ref.~\citen{Feneberg2015}

Experimental data of both samples were modeled in an anisotropic two layer model consisting of the bulk V$_2$O$_5$ and an anisotropic Bruggeman\cite{Bruggeman1935} effective medium approximated surface roughness consisting of 50$\%$ oxide and 50$\%$ void. A model independent point-by-point fit to the experimental data yielded the complex elements of the dielectric tensor $\varepsilon = \varepsilon_1 + i \varepsilon_2$. It is important to take the uniaxial structure of the material correctly into account to describe the peak amplitudes correctly. Due to the surface orientation of the samples, components $\varepsilon_x$ and $\varepsilon_y$ are determined with high accuracy while $\varepsilon_z$ has higher error bars. Nevertheless, the amplitude of $\varepsilon_{2z}$ does not exceed 3 in the investigated spectral range.

\section*{Data availability}
The data that support this work are available in the article and Supplementary information file.

\section*{Acknowledgements}
This work benefited from the support of EDF in the framework of the research and teaching Chair ``Sustainable energies'' at Ecole Polytechnique. Computational time was granted by GENCI (Project No. 544).
WRLL was supported by the U.S. Department of Energy - Basic Energy Sciences (DOE-BES) grant no. DE-SC0008933. This material is in part based upon work supported by the National Science Foundation under grant no. DMR-1555153.
This research is part of the Blue Waters sustained-petascale computing project, which is supported by the National Science Foundation (awards OCI-0725070 and ACI-1238993) and the state of Illinois.
Blue Waters is a joint effort of the University of Illinois at Urbana-Champaign and its National Center for Supercomputing Applications. The experimental part of this work was performed in the framework of GraFOx, a Leibniz-Science Campus
partially funded by the Leibniz association.  We thank Carsten Hucho and Arno Wirsig (Paul-Drude-Institut f\"{u}r Festk\"{o}rperelektronik, Leibniz-Institut im Forschungsverbund Berlin e.V., Germany) for the generous access to their samples.

\section*{Author contributions}

V.G., L.R., W.R.L.L. and M.G. designed research; A.S. did the first preliminary BSE calculations
; V.G. performed all the fully first principles QSGW and BSE calculations; M.F. conducted the experiments; V.G., L.R., and M.G analysed the excitons and proposed the model; V.G., L.R., W.R.L.L. and M.G. wrote the manuscript with contributions from all authors. All authors reviewed the manuscript.

\section*{Competing interests}

The authors declare no competing interests.


\section*{Figure legends}

Figure 1. \textbf{Absorption spectra of \vo for $x$, $y$ and $z$ polarization}  (the \vo  layers are in the $xy$ plane). The experimental spectra obtained from ellipsometry  measurements  are compared with calculated BSE spectra in the three directions (note the change of intensity scale for the $z$ direction in the right panel). The vertical arrows mark the position of the dark excitons below the absorption onset.

Figure 2. \textbf{Electron density distribution for dark and bright excitons.} (a) Crystal structure of \vo~. V atoms are grey, while O atoms are red. Selected atoms indicate the V-O-V rungs. (b)-(c) Electron density distribution $|\Psi_\lambda(\bfr_h^0,\bfr_e)|^2$ for the first exciton $\lambda$ that is bright in $x$ direction and (d)-(e) for the lowest-energy dark exciton. In both cases, the position of the hole $\bfr_h^0$ is fixed close to a bridge oxygen O$_b$ atom (light blue dot). Since it is displaced along a non-symmetric direction from the O$_b$ atom the resulting electronic distribution is also slightly asymmetric. (b) and (d) views in the plane of \vo layers (i.e., $xy$ planes) ; (c) and (e) views normal to \vo layers (i.e., $zy$ planes). Ellipses centered at the hole position represent the calculated exciton mean radius $\langle |r^{\alpha}_{\lambda}| \rangle$. The isosurface value for bright and dark excitons is chosen to be 1\% of the maximum electron density.

Figure 3. \textbf{Analysis of the absorption spectra of \vo for x, y, and z polarization.} The calculated JDOS and independent-particle (IPA) spectra  from the QSGW band structure are compared with the RPA spectra that include local field effects (LFE) and the BSE spectra that also include excitonic effects. The excitonic JDOS from the BSE, which is the density of excitonic states at zero momentum as a function of energy, is also shown for comparison. Discrete bound excitonic states inside the gap give rise to a JDOS with small intensity, not visible at the scale of the plot. The vertical arrows mark the position of the dark excitons below the absorption onset.

Figure 4.
\textbf{Contributions to the optical absorption intensity.} 
(a) For first bright excitons contributing to the main peaks in $x$ and $y$ directions (at 3.06 and 3.05 eV respectively). Size of the circles indicates $| \sum_{vc \kvec} \bar{A}^{vc \kvec}_{\lambda} \tilde{\rho}_{vc \kvec} |$ summed over equivalent $\kvec$ values. 
(b) For bright and dark exciton in $x$ direction. Size of the circles indicates $|\bar{A}^{vc \kvec}_{\lambda} \tilde{\rho}_{vc \kvec} |$  In both (a) and (b), transitions are vertical, so each transition links two circles of equal size, one in a valence and one in a conduction band, at the same ${\bf k}$ point.

\end{document}